\documentstyle[prb,aps,preprint,epsfig]{revtex}
\begin{document}
\draft

\title{TUNNELING WITH DISSIPATION IN OPEN QUANTUM SYSTEMS}
\author{ G.G.Adamian$^{1,2}$, N.V.Antonenko$^{1,2}$ and
W.Scheid$^{1}$}
\address{$^{1}$Institut f\"ur Theoretische Physik der
Justus--Liebig--Universit\"at,
D--35392 Giessen, Germany\\
$^{2}$Joint Institute for Nuclear Research, 141980 Dubna, Russia}
\date{\today}
\maketitle

\begin{abstract}
Based on the general form of the master equation
for open quantum systems the tunneling
is considered. Using the path integral technique a simple closed
form expression for the tunneling rate through a parabolic barrier
is obtained. The tunneling in the open quantum systems
strongly depends on the coupling with environment.
We found the cases when the dissipation prohibits
tunneling through the barrier but
decreases the crossing of the barrier for the energies
above the barrier. As a particular application, the case of
decay from the metastable state is considered.
\end{abstract}


%{\bf 1. Introduction}\\[3mm]
There has been considerable interest to the quantum
tunneling of a particle through an energy
barrier when the dissipation is present
\cite{1,2,3,4,5,6,7,8,9,10,11,12,13,14,15}.
Using various models for the description  of
the quantum open system, the opposite dependences of tunneling
rate on the dissipation have been observed.
It is generally thought that tunneling probability
decreases in the presence of coupling to the environment.
Disregarding the stage of averaging over the intrinsic degrees of freedom,
one can consider the tunneling effect
starting right away from the general Markovian master equation
for the reduced density matrix  of the collective degree of freedom
\cite{16,17,18,19,20,21,22,23,24,25}
\begin{eqnarray}
\frac{d\hat\rho (t)}{dt}=-\frac{i}{\hbar}[\hat H_0,\hat\rho]+
\frac{1}{2\hbar}\sum_{j}\left([\hat V_j\hat\rho, \hat V_j^+]+
[\hat V_j, \hat\rho \hat V_j^+]\right).
\label{1_eq}
\end{eqnarray}
Here, $\hat H_0$ is the Hamiltonian of the collective subsystem and
$\hat V_j$
are operators acting in the Hilbert space of the subsystem.
The second term in (\ref{1_eq}) is responsible for the friction and
diffusion and supplies the unreversability in the open quantum system.
Omitting this term we get a standard form for the density
matrix evolution equation in the case of closed system.
The generality of Eq. (\ref{1_eq}) was mathematically
proved in \cite{18,19}.

In the one-dimensional case the phase space path integral
expression \cite{26}
for the propagator corresponding to (\ref{1_eq})
is written as
\begin{eqnarray}
G(q,q',t;q_0,q^{'}_0,0)&=&\int\limits_{q_0(0)}^{q(t)}D[\alpha]
\int\limits_{q^{'}_0(0)}^{q'(t)}D[\alpha']
\exp{(\frac{i}{\hbar}S[\alpha, \alpha'])}, \nonumber \\
S[\alpha, \alpha']&=&
\int\limits_{0}^{t}d\tau \{\dot q(\tau)p(\tau) - 
H_{eff}(q(\tau),p(\tau))\}
\nonumber \\
&-&
\int\limits_{0}^{t}d\tau \{\dot q'(\tau)p'(\tau) -
H^{*}_{eff}(q'(\tau),p'(\tau))\}
\nonumber \\
&-&i\sum_{j}^{}\int\limits_{0}^{t}d\tau \{V_j(q(\tau),p(\tau))
 V^{*}_j(q'(\tau),p'(\tau))\},
\label{2_eq}
\end{eqnarray}
with the effective Hamiltonian
$$H_{eff}=H_0 - \frac{i}{2}\sum_{j}^{}|V_j|^2.$$
Here, the Wigner transform of operators $\hat H_0$
$\hat V^+_j\hat V_j$, $\hat V^+_j$ and
$\hat V_j$ are denoted by $H_0$, $|V_j|^2$, $V_j^*$ and $V_j$,
respectively.
For the inverted harmonic oscillator with the Hamiltonian
\begin{eqnarray}
\hat H_0=\frac{1}{2m}\hat
p^2-\frac{m\omega^2}{2}\hat q^2
\label{3_eq}
\end{eqnarray}
and linear environment operators
\begin{eqnarray}
\hat V_j=A_j\hat p +B_j\hat q,\quad
\hat V_j^+=A_j^*\hat p +B_j^*\hat q, \quad j=1,2,
\label{4_eq}
\end{eqnarray}
the propagator (\ref{2_eq}) can be evaluated analytically:
\begin{eqnarray}
G(q,q',t;q_0,q^{'}_0,0)=\frac{m\omega}{2\pi \hbar \sinh(\omega t)}
\exp (\lambda t)\exp(iS_R/\hbar)
\exp(-S_I/\hbar),
\label{5_eq}
\end{eqnarray}
where
\begin{eqnarray}
S_R&=&\frac{m\omega}{2\sinh(\omega t)}
(\cosh(\omega t)[q_0^2 - (q^{'}_0)^2+q^2-(q^{'})^2] \nonumber\\
&-& 2\cosh(\lambda t)[q_0q-q_0^{'}q']
-2\sinh(\lambda t)[q_0q'-q_0^{'}q]),\nonumber
\end{eqnarray}
$$S_I=\frac{m}{8\lambda(\omega^2-\lambda^2)
\sinh^2(\omega t)}({\cal A}(t)[q_0-q_0^{'}]^2-
4{\cal B}(t)[q_0-q_0^{'}][q-q^{'}]
-{\cal A}(-t)[q-q^{'}]^2),$$
\begin{eqnarray}
{\cal A}(t)&=&a\exp(2\lambda t)
+b\sinh(2\omega t) + c\cosh(2\omega t) -d,\nonumber\\
{\cal B}(t)&=&
a\cosh(\omega t)\sinh(\lambda t) +
b \sinh(\omega t)\cosh(\lambda t),\nonumber\\
a&=&\frac{2\omega^2}{\hbar m}(m^2D_{qq}[\omega^2-2\lambda^2]-
2\lambda mD_{pq} - D_{pp}),\nonumber\\
b&=& \frac{2\omega\lambda}{\hbar m}(m^2\omega^2D_{qq}
+2\lambda mD_{pq}+ D_{pp}),\nonumber\\
c&=& \frac{2\lambda}{\hbar m}(m^2\omega^2\lambda D_{qq}
+\lambda D_{pp}+ 2m\omega^2D_{pq}),\nonumber\\
d&=& \frac{2}{\hbar m}(\omega^2-\lambda^2)
(m^2\omega^2 D_{qq} -D_{pp}).\nonumber
\end{eqnarray}
Here, the quantum mechanical diffusion coefficients
$D_{qq}=\frac{\hbar}{2}\sum\limits_{j}|A_j|^2$,
$D_{pp}=\frac{\hbar}{2}\sum\limits_{j}|B_j|^2$ and
$D_{qp}=-\frac{\hbar}{2}{\rm Re}\sum\limits_{j}A_j^*B_j$
and the frictional damping rate
$\lambda=-{\rm Im}\sum\limits_{j}A_j^*B_j$
\cite{18,19,22,23,24,25} satisfy the following constraints:
$D_{qq}>0$, $D_{pp}>0$ and $D_{pp}D_{qq}-D_{pq}^2\ge \lambda^2\hbar^2/4$
which secure the non-negativity of the density matrix at any time.
The diffusion models, in which these constraints are not fulfilled,
can be related to the classical or semiclassical considerations
because they allow the violation of the uncertainty inequality
at some time \cite{15,20,21,22,23,24}.

Using (\ref{5_eq}),  $\hat\rho (t)$
is determined from  $\hat\rho (t=0)$ as
\begin{eqnarray}
<q|\hat\rho (t)|q'>=\int\limits_{}^{}dq_0\int\limits_{}{}dq^{'}_0
G(q,q',t;q_0,q^{'}_0,0)<q_0|\hat\rho(t=0)|q^{'}_0>.
\label{6_eq}
\end{eqnarray}
In order to study the
tunneling with the Hamiltonian (\ref{3_eq}),
we consider a particle in the initial state
\begin{eqnarray}
\Psi(q)=  (2\pi \sigma_{qq}(0))^{-1/4}
\exp(-\frac{1}{4\sigma_{qq}(0)}(q -\bar q(0))^2 + \frac{i}{\hbar}\bar p(0)q)
\label{7_eq}
\end{eqnarray}
in the left-hand side from the potential barrier.
The calculation of (\ref{6_eq}) with (\ref{5_eq}) and (\ref{7_eq})
yields the Gaussian distribution at time $t$
\begin{eqnarray}
\rho (q,t)=<q|\hat\rho (t)|q>=
(2\pi \sigma_{qq}(t))^{-1/2}
\exp(-\frac{1}{2\sigma_{qq}(t)}(q -\bar q(t))^2),
\label{8_eq}
\end{eqnarray}
with the first $\bar q(t)$ and second $\sigma_{qq}(t)$ moments.
The equations for these moments are given in
Refs.~\cite{15,18,19,21,22,23,24} and below for
arbitrary potential.
Originally they contain the friction in both coordinate $\lambda_q$
and momentum $\lambda_p$ so that $\lambda_p+\lambda_q=2\lambda$.
The considered particular case of $\lambda_p=\lambda_q=\lambda$
is generalized for $\lambda_p\ne \lambda_q$ by using the
canonical transformations \cite{15}
$p'=p+\mu mq$, $q'=q$ with the parameter $\mu$.
Therefore, the expression
(\ref{8_eq}) can be applied to the case of
$\lambda_p\ne \lambda_q$ as well.

The solutions of equations for the first and second moments in
(\ref{8_eq}) are
\begin{eqnarray}
\bar q(t)&=&
e^{-\lambda t}\left(\bar q(0)\left[\cosh(\psi t/2)+
\frac{\lambda_p-\lambda_q}{\psi}\sinh(\psi t/2)\right]+
\frac{2}{m\psi}\bar p(0)\sinh(\psi t/2)\right),\nonumber\\
\sigma_{qq}(t)&=&
\frac{1}{2m^2\lambda(\omega^2-\lambda_p\lambda_q)}
\left[m^2(\omega^2-2\lambda_p\lambda)D_{qq}-D_{pp}-2m\lambda_pD_{pq}\right]
\nonumber\\
&+& e^{-2\lambda t}\left[
\frac{2C_1}{m(\lambda_q-\lambda_p)}-
\frac{1}{2m\omega^2}[(\lambda_q-\lambda_p)C_2+C_3\psi]\cosh(\psi t)\right.
\nonumber\\
&+&\left.\frac{1}{2m\omega^2}[(\lambda_q-\lambda_p)C_3+C_2\psi]\sinh(\psi t)
\right],
\label{10a_eq}
\end{eqnarray}
where the following notations are used:
\begin{eqnarray}
C_1&=&\frac{m\omega^2(\lambda_q-\lambda_p)}{\psi^2}\left[
\sigma_{qq}(0)-\frac{1}{m^2\omega^2}\sigma_{pp}(0)+
\frac{\lambda_q-\lambda_p}{m\omega^2}\sigma_{pq}(0)\right.\nonumber\\
&-&\left.\frac{1}{\lambda}D_{qq}+\frac{1}{m^2\omega^2\lambda}D_{pp}-
\frac{(\lambda_q-\lambda_p)}{m\omega^2\lambda}D_{pq}\right],\nonumber\\
C_2&=&\frac{1}{\psi^2}\left[
\frac{\lambda_q-\lambda_p}{m}(\sigma_{pp}(0)-
m^2\omega^2\sigma_{qq}(0))+
4\omega^2\sigma_{pq}(0)\right.\nonumber\\
&+&\left.
\frac{1}{\omega^2-\lambda_p\lambda_q}
(\frac{2\omega^2-\lambda_p\lambda_q+\lambda_q^2}
{m}[D_{pp}+
m^2\omega^2D_{qq}]+
4\lambda\omega^2D_{pq})\right],\nonumber\\
C_3&=&-\frac{1}{m\psi}\left[
m^2\omega^2\sigma_{qq}(0)+
\sigma_{pp}(0)\right.\nonumber\\
&+&\left.\frac{1}{\omega^2-\lambda_p\lambda_q}(
\lambda_qD_{pp}+ 2m\omega^2D_{pq}+
m^2\omega^2\lambda_pD_{qq})\right]\nonumber
\end{eqnarray}
and $\psi =\sqrt{(\lambda_p-\lambda_q)^2 + 4\omega^2}$.
With these expressions
we obtain the same result as in Ref.~\cite{27}
at $\lambda_p=\lambda_q=0$, $D_{pp}=D_{qq}=D_{pq}=0$
and $\sigma_{pp}(0)=\hbar^2/(4\sigma_{qq}(0))$
($\sigma_{qp}(0)=0$).
For $\lambda_q=0$, $D_{pp}=D_{qq}=D_{pq}=0$ and
$\sigma_{pp}(0)=\hbar^2/(4\sigma_{qq}(0))$
($\sigma_{qp}(0)=0$), our results coincide
in the underdamped limit
with the results of Ref.~\cite{4}
where the tunneling was studied with
the inverted Caldirola-Kanai Hamiltonian.

The penetration probability at time $t$ is determined by the following
expression ($q=0$ corresponds to the top of the barrier):
\begin{eqnarray}
P(t)=\int\limits_{0}^{\infty}dq[\rho (q,t)-\rho (q,t=0)]/
\int\limits_{-\infty}^{0}dq\rho (q,0),
\label{10_eq}
\end{eqnarray}
which is the ratio of change of the probability to be
on the right-hand side of the barrier in time $t$
over the initial probability
of the finding the particle on the entry left-hand side.
Using Eqs. (\ref{8_eq})-(\ref{10_eq}), the penetration
probability $P=P(t\to\infty)$ is easially calculated
taking the initial variances
in accordance with the uncertainty relation.
Here, we use
$\sigma_{qq}(0)\sigma_{pp}(0)=\hbar^2/4$ and $\sigma_{pq}(0)=0$.

The dependences of the penetration probability through the
parabolic barrier on the initial energy $E$ of system are
presented in Fig.~1  for three sets of the friction
coefficients $\lambda_p$ and $\lambda_q$.
All diffusion coefficients depend only on $\lambda$.
For the sub-barrier energies ($E<0$), the tunneling
is larger for $\lambda_q=\lambda_p\ne 0$ as comparable
to the case without friction $\lambda_p=\lambda_q=0$.
For $E<0$, the dissipation in coordinate
$\lambda_q$ increases but dissipation in momentum $\lambda_p$
decreases the barrier penetration.
The increase of the tunneling was obtained
in the microscopic Gisin's model \cite{28} for large friction.
However, in this model one can not distinguish the
influence of frictions in coordinate
and momentum on the tunneling.
Larger penetration of the barrier than in the standard
coupled-channel calculations is necessary to explain the
experimental data on the sub-barrier fusion ~\cite{29}.
It could be that in this case the coupling with environment
leads to $\lambda_q\ne 0$ that renormalizes the barrier
and increases the penetration \cite{15}.
The friction and diffusion reduce the crossing of the barrier
for the energies above the barrier.
For $E=0$ and $\lambda_p=\lambda_q$, the penetration
and reflection probabilities
are equal to each other with and without dissipation.

In Fig.~2 we show how the tunneling depends on the
diffusion coefficients at different values of friction
in momentum $\lambda_p$ for $\lambda_q=0$. One can see that only
with the diffusion coefficient in momentum $D_{pp}$
($D_{qq}=D_{pq}=0$) $P$ decreases with increasing
$\lambda_p$. Note that this set of diffusion
coefficients is not compatible with the quantum mechanical
consideration.
For $D_{qq}\ne 0$, the value of  $P$ initially decreases with
increasing $\lambda_p$ up to some "critical" friction coefficients
and then it starts to grow.
This effect becomes more evident at larger
$D_{qq}$ and $D_{pp}$ (higher temperature).
The  "critical" friction coefficient decreases with increasing
temperature.
This behaviour of the tunneling probability
$P$ as a function of $\lambda_p$ can be explained in the following way:
The tunneling is more crucial to the value of
$D_{qq}$ then to the value of $D_{pp}$ because $\sigma_{qq}(t)$
(correspondingly $\rho(q,t)$ and $P$) is more sensitive to
$D_{qq}$ than to $D_{pp}$; At large $\lambda_p$ the
system has a longer time for the tunneling and during this time
$\sigma_{qq}(t)$ and $P(t)$ strongly increase due to diffusion
in coordinate. The increase of tunneling rate with temperature
is in agreement with Ref.~\cite{3}.

The probability of finding the particle to the right of the barrier
is very sensitive to the width $\sigma_{qq}(0)$
of the initial wave packet localized
to the left of the barrier at $t=0$ (Fig.~3).
This effect is weaker with the dissipation.
For smaller $\sigma_{qq}(0)$, the value of $\sigma_{pp}(0)$
becomes larger in quantum mechanics and the penetration probability
increases due to the larger fluctuation energy.
In the vicinity of $\sigma_{qq}(0)=\hbar/(2m\omega)$ the dependence
of $P$ on $\sigma_{qq}(0)$ becomes weak and
the curve in fig.3 has a step-like behaviour.

The calculated time of decay from the metastable state
in the potential
\begin{eqnarray}
U(q)=\alpha q^2-\beta q^3
\label{11_eq}
\end{eqnarray}
is shown in Fig.~4 as a function of $\lambda_p$.
These data result from the solution of equations
for the first and second moments (obtained from eq. (\ref{1_eq})):
\begin{eqnarray}
\frac{d\bar q}{dt}&=&-\lambda_q\bar q +\frac{1}{m}\bar p,\nonumber\\
\frac{d\bar p}{dt}&=&-\partial {U(\bar q)}/{\partial {\bar q}}-
\frac{1}{2}\partial^3 {U(\bar q)}/{\partial {\bar q}^3}\sigma_{qq} -
\lambda_p\bar p,
\nonumber\\
\frac{d\sigma_{qq}}{dt}&=&-2\lambda_q\sigma_{qq}+
\frac{2}{m}\sigma_{pq}+2D_{qq},\nonumber\\
\frac{d\sigma_{pp}}{dt}&=&-2\lambda_p\sigma_{pp}-
2\partial^2 {U(\bar q)}/{\partial {\bar q}^2}\sigma_{pq}+2D_{pp},\nonumber\\
\frac{d\sigma_{pq}}{dt}&=&-\partial^2 {U(\bar q)}/{\partial
{\bar  q}^2}\sigma_{qq}+
\frac{1}{m}\sigma_{pp}-(\lambda_p+\lambda_q)\sigma_{pq}+2D_{pq}.
\label{10b_eq}
\end{eqnarray}
These equations are obtained from (\ref{1_eq}) and (\ref{4_eq})
for arbitrary potential $U(q)$. In order to calculate $P(t)$
for short times, we can use in the first approximation
the formalism elaborated for the parabolic barrier.
The value of time $t_{1/2}$ at which $P(t_{1/2})=0.5$
(the value of $\bar q$ corresponds to the
top of the barrier)
may be defined in some sense as the tunneling time
\cite{7}. The tunneling time increases monotonically
with $\lambda_p$ ($\lambda_q=0$) when $D_{qq}=0$. For
$D_{qq}\ne 0$, the value of $t_{1/2}$ initially increases with
$\lambda_p$ and then it decreases.
This means that for large $\lambda_p$
the dissipation prohibits the decay from metastable state
due to diffusion in coordinate.
The results of calculations above
are in agreement with the results obtained in \cite{7}
using the Gisin model \cite{28} for the double well potential.

In conclusion,
our calculations show that the dissipative effects on the
tunneling process are quite complicated. It is evident
that the earlier conclusions that the dissipation inhibit
tunneling is not correct in the general case.
There are examples when the dissipation prohibits the
penetration through the barrier. Using the general master
equation (\ref{1_eq}) for describing
the open quantum systems, we can transparently show
the influence of each friction and diffusion coefficient
on the tunneling. However, the microscopical
calculation of these coefficients in the real
system remains to be interesting problem. In the consistent
quantum treatment the tunneling should be calculated
with the set of the diffusion coefficients where
$D_{qq}>0$. As was shown, the tunneling is crucial
to the value of $D_{qq}>0$. If the environment operators
lead to $\lambda_q\ne 0$ then the interaction with
environment renormalizes the potential barrier and
influence the tunneling. With the initial Gaussian
distribution (\ref{7_eq}) the distribution function
remains to be Gaussian at any time.

The author (N.V.A.) is grateful to the Alexander
von Humboldt-Stiftung for the financial support. This work was
supported in part by DFG.

\begin{figure}
\caption{Calculated dependence of the
penetration probability through the parabolic barrier
on the initial energy of particle $E$ at temperature
$kT=0$, $\hbar\omega=2.0$ MeV, $q(0)=-2$ fm,
$\sigma_{qq}(0)=0.2$ fm$^2$,
$m=53m_0$ ($m_0$ is the mass of nucleon),
$D_{qq}=\hbar\lambda/(2m\omega)$,
$D_{pp}=\lambda m\hbar\omega/2$ and $D_{pq}=0$.
The results for the cases
($\lambda_p=\lambda_q=0$), ($\hbar\lambda_p=\hbar\lambda_q=1$ MeV)
and ($\hbar\lambda_p=2$ MeV, $\lambda_q=0$)
are presented by solid, dotted and dashed lines,
respectively.}
\end{figure}

\begin{figure}
\caption{Calculated dependence of the
tunneling probability
$P$ on the friction coefficient in momentum $\lambda_p$ at
temperatures $kT=0$ and 3 MeV, $\hbar\omega=2.0$ MeV,
$q(0)=-2$ fm, $\sigma_{qq}(0)=0.2$ fm$^2$, $E=-5$ MeV, $m=53m_0$,
and friction coeffisients in coordinate $\lambda_q=0$
($\lambda=\lambda_p/2$).
The calculations for the cases
($kT=0$, $D_{qq}=\hbar\lambda/(2m\omega)$,
$D_{pp}=\lambda m\hbar\omega/2$ and $D_{pq}=0$),
($kT=0$, $D_{qq}=0$, $D_{pp}=\lambda_pm\hbar\omega/2$ and
$D_{pq}=0$),
($kT=3$ MeV, $D_{qq}=\hbar\lambda/(2m\omega)
\coth(\hbar\omega/(2kT))$,
$D_{pp}=\lambda m\hbar\omega/2 \coth(\hbar\omega/(2kT))$ and
$D_{pq}=0$) and
($kT=3$ MeV, $D_{qq}=0$,
$D_{pp}=\lambda_pm\hbar\omega/2\coth(\hbar\omega/(2kT))$ and
$D_{pq}=0$) are presented by
solid, dashed, dotted and dashed--dotted lines, respectively.}
\end{figure}

\begin{figure}
\caption{Calculated dependence of the
penetration probability
$P$ on the initial variance $\sigma_{qq}(0)$ at
$kT=0$, $\hbar\omega=2.0$ MeV,
$q(0)=-1$ fm, $p(0)=0$, $m=53m_0$,
$D_{qq}=\hbar\lambda/(2m\omega)$,
$D_{pp}=\lambda m\hbar\omega/2$ and $D_{pq}=0$.
The results obtained with $\lambda=\lambda_p=\lambda_q=0$
and $\hbar\lambda=\hbar\lambda_p=\hbar\lambda_q=1$ MeV
are presented by solid and dashed lines, respectively.
The value $\sigma_{qq}(0)=\hbar/(2m\omega)$ is marked by arrow.}
\end{figure}

\begin{figure}
\caption{Calculated dependence of the
decay time from the metastable
state in the potential (\protect\ref{11_eq}) on the friction coefficient
$\lambda_p$ at $\lambda_q=0$ ($\lambda=\lambda_p/2$), $kT=0$,
$p(0)=0$, $\sigma_{qq}(0)=0.2$ fm$^2$ and $m=53m_0$.
The depth of potential pocket with the minimum at
$q(0)=-1.08$ fm is 5 MeV ($\alpha=-2.57$ MeV fm$^{-2}$ and
$\beta=1.59$ MeV fm$^{-3}$).
The top of the barrier corresponds
$E=0$ MeV at $q=0$ fm.
The calculations for the cases
($D_{qq}=\hbar\lambda/(2m\omega)$,
$D_{pp}=\lambda m\hbar\omega/2$ and $D_{pq}=0$) and
($D_{qq}=0$, $D_{pp}=\lambda_pm\hbar\omega/2$ and
$D_{pq}=0$)
are presented by solid and dotted lines,
respectively.}
\end{figure}

\newpage
\epsfig{figure=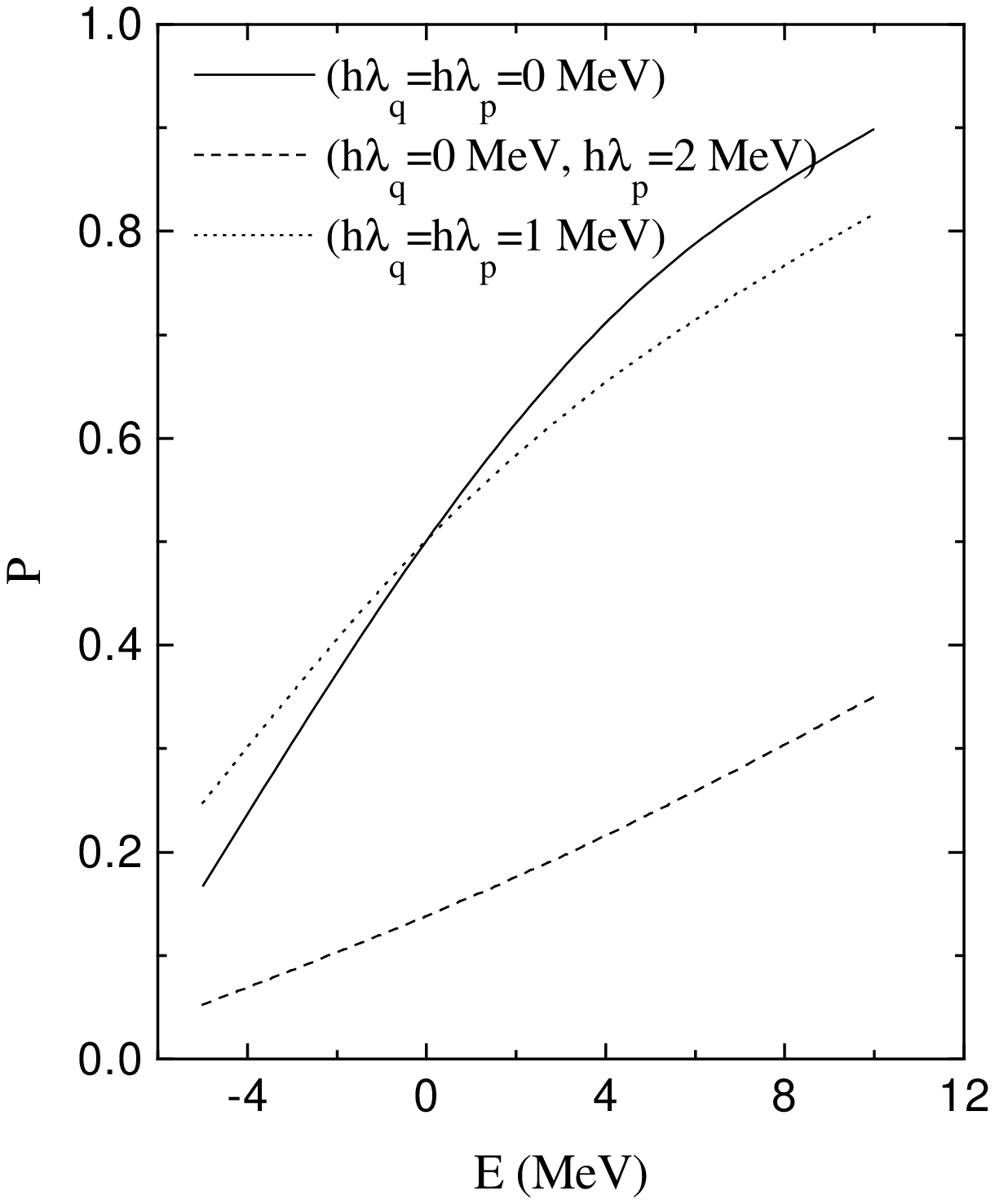,width=16cm,height=23cm}
\epsfig{figure=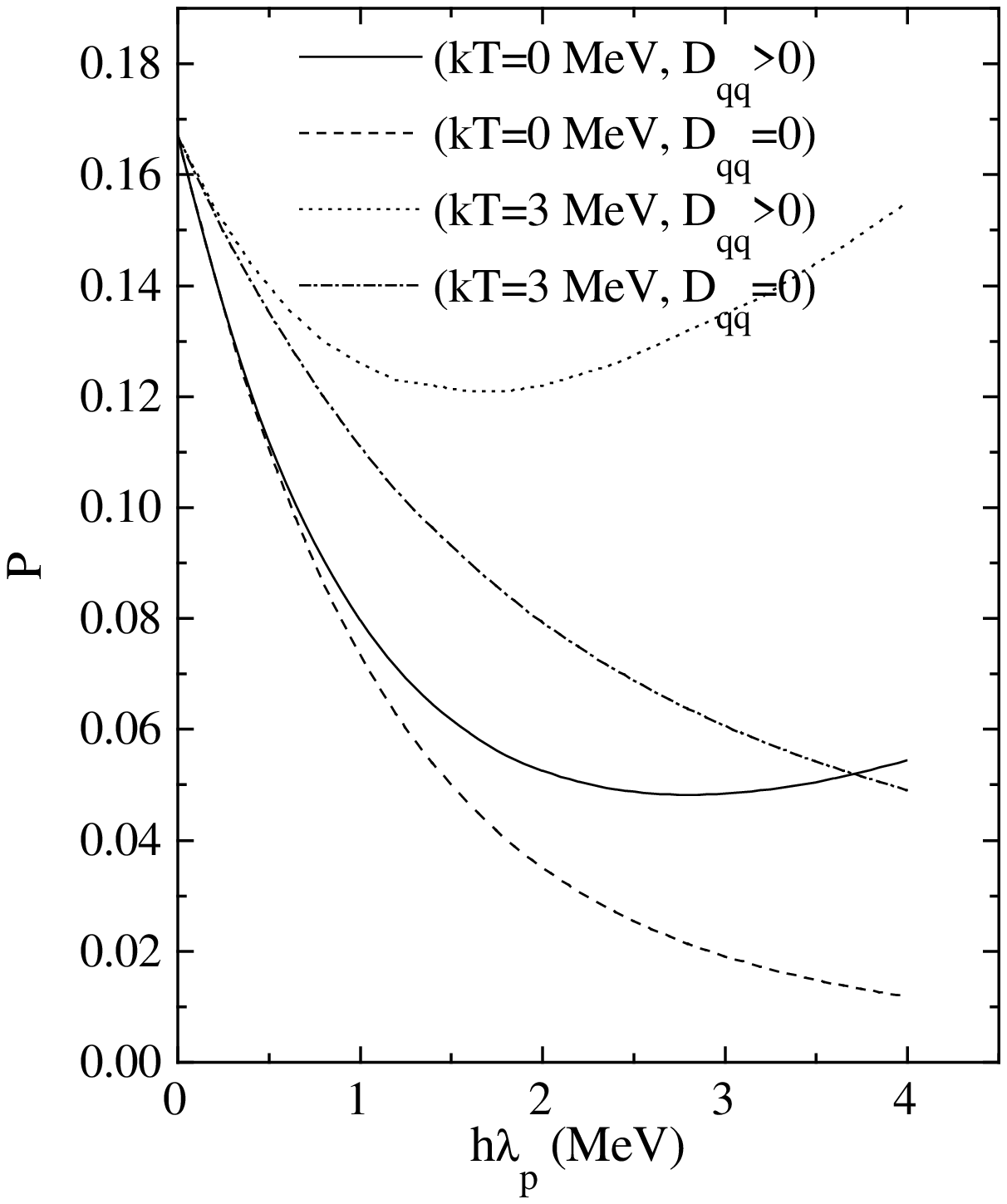,width=16cm,height=23cm}
\epsfig{figure=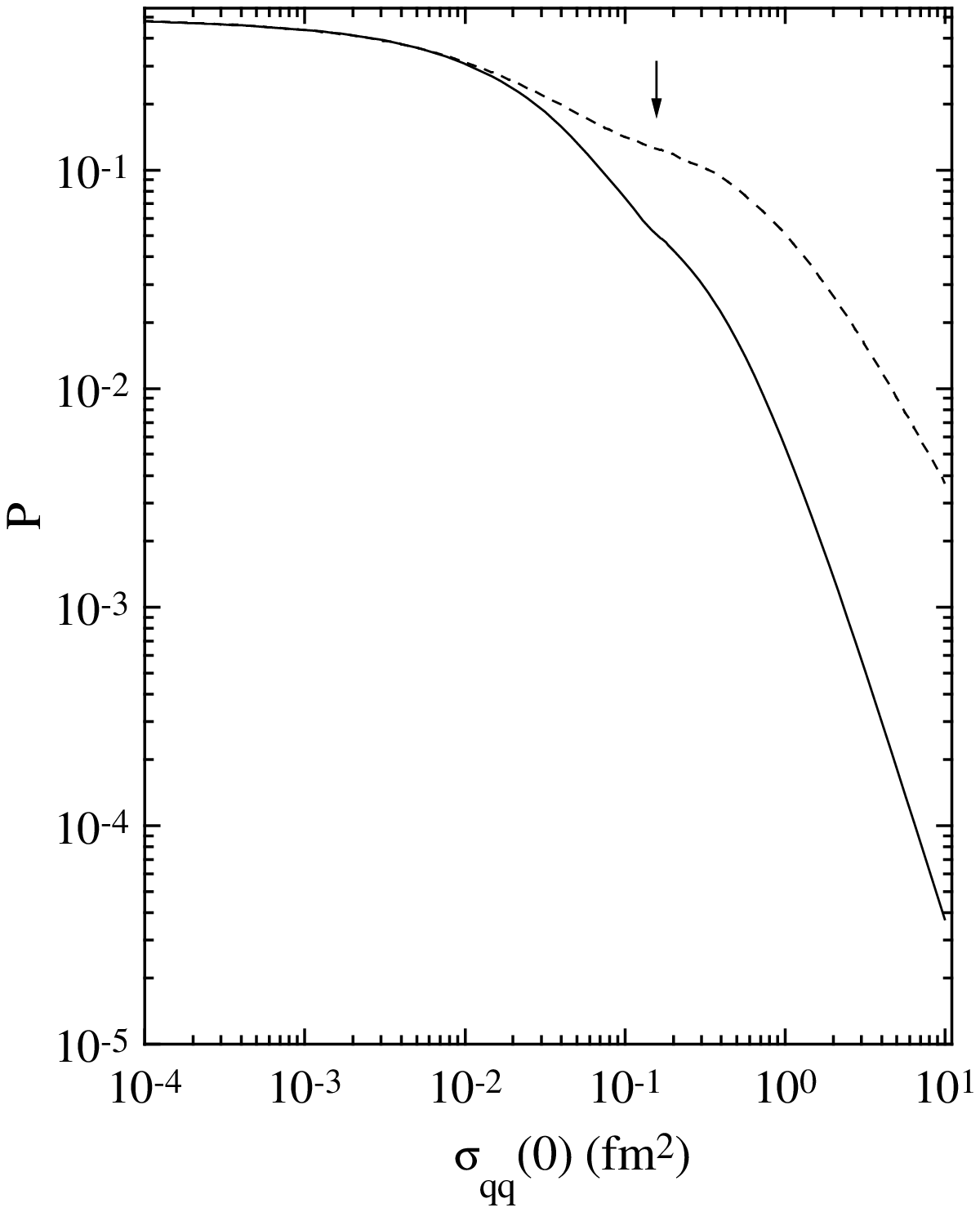,width=16cm,height=23cm}
\epsfig{figure=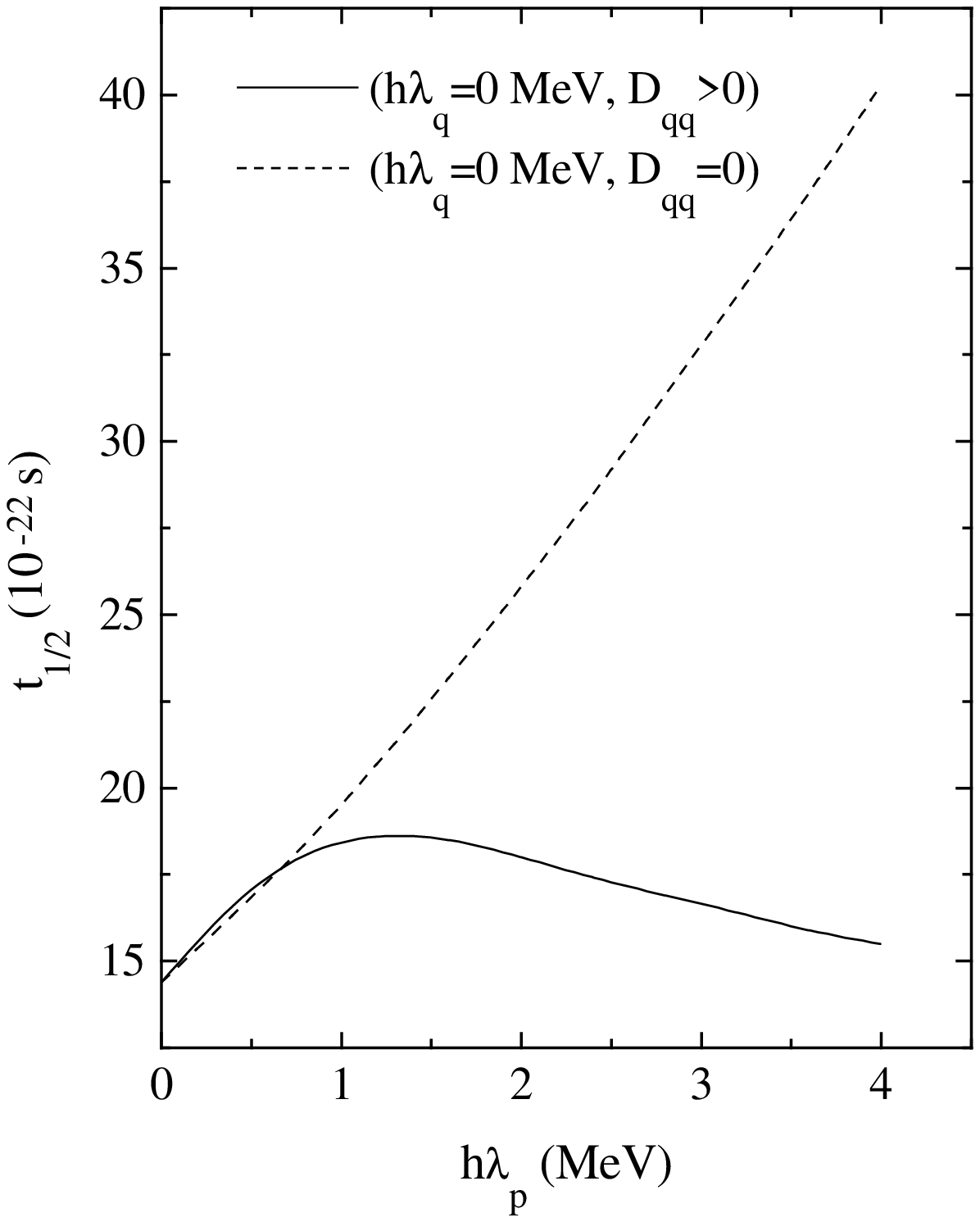,width=16cm,height=23cm}

\end{document}